\documentstyle[12pt]{article}

\def\void{}
\def\labelmark{}

\newenvironment{formula}[1]{\def\labelname{#1}
\ifx\void\labelname\def\junk{\begin{displaymath}}
\else\def\junk{\begin{equation}\label{\labelname}}\fi\junk}%
{\ifx\void\labelname\def\junk{\end{displaymath}}
\else\def\junk{\end{equation}}\fi\junk\labelmark\def\labelname{}}

\newcommand{\beq}{\begin{formula}}
\newcommand{\eeq}{\end{formula}}
\newcommand{\beqv}{\begin{formula}{}}

\newcommand{\bea}{\begin{eqnarray}}
\newcommand{\eea}{\end{eqnarray}}
\newcommand{\rf}[1]{(\ref{#1})}

\renewcommand{\b}{\beta}
\renewcommand{\a}{\alpha}

\newcommand{\k}{\kappa}

\newcommand{\del}{\delta}

\newcommand{\cN}{{\cal N}}

\newcommand{\noi}{\noindent}

\begin{document}
\topmargin 0pt
\oddsidemargin 5mm
\headheight 0pt
\topskip 0mm

\addtolength{\baselineskip}{0.20\baselineskip}

\hfill NBI-HE-94-29

\hfill May 1994
\begin{center}

\vspace{36pt}
{\large \bf
On the exponential bound \\
in four dimensional simplicial gravity}

\vspace{24pt}

{\sl J. Ambj\o rn } and {\sl J. Jurkiewicz}\footnote{Permanent address:
Institute of Physics, Jagellonian University, ul. Reymonta 4, PL-30 059,
Krakow 16, Poland}

\vspace{12pt}

The Niels Bohr Institute\\
Blegdamsvej 17, DK-2100 Copenhagen \O , Denmark\\

\end{center}

\vfill

\begin{center}
{\bf Abstract}
\end{center}

\vspace{12pt}

\noi
Simplicial quantum gravity has been proposed as a regularization for
four dimensional quantum gravity. The partition function is constructed
by performing a weighted sum over all triangulations of the 4-sphere.
The model is well-defined only if the number of such triangulations
consisting of $N$ simplexes is exponentially bounded. Numerical simulations
seem so far to favor such a bound.
\vspace{24pt}

\vfill

\newpage

\noi{\large \bf 1. Introduction}

\vspace{12pt}

\noi
"Simplicial quantum gravity" has been proposed as
a regularization for (Euclidean) quantum gravity in four dimensions
\cite{aj,am}. It was suggested as a natural generalization of similar
models studied in two dimensions \cite{david1,adf,kkm}
and three dimensions \cite{adj,av,sakura,gross,am1,bk,av1,abkv}.
In all these models the partition functions are given by
\beq{*1}
Z = \sum_T \frac{1}{C_T}e^{-S(T)}
\eeq
where the summation is over all combinatorially inequivalent triangulations
of a manifold of {\it fixed} topology. The factor $1/C_T$ is the symmetry
factor of the triangulation. It is not important for the
following discussion. For pure gravity $S(T)$ is bounded by
\beq{*2}
|S(T)| < \tilde{c} |T|
\eeq
where $|T|$ denotes the number of simplexes in the triangulation $|T|$.
The inequality \rf{*2} implies that the partition function is well defined only
if the
number $\cN (|T|)$ of triangulations of a fixed topology (which we
in the following  will assume spherical) and consisting of $|T|$ simplexes
is exponentially bounded:
\beq{*3}
\cN(|T|) \leq e^{c |T|}.
\eeq
In the case of $2d$ simplicial quantum gravity the exponential bound was
proven long ago \cite{tutte}. In $3d$ a number of plausible arguments was given
in favor of an exponential bound \cite{adj},
but no  complete proof exists so far.
Most of the arguments of \cite{adj} extend to $4d$, but
again a proof is missing. In the absence of a proof
in $3d$ the question was addressed numerically in \cite{av}, and good
evidence in favor of an exponential bound was found.
The situation in $4d$ seems very similar to the situation in
$3d$ and the finite size effects associated with the identification of the
smallest possible $c$ in \rf{*3} were even considerably smaller than
than in $3d$. Recently it has been claimed, however, based on new
numerical simulations,
that  the situation in $4d$ differs profoundly from the situation in
$2d$ and $3d$ and that $\cN(|T|)$ grows factorially \cite{ckr}:
\beq{*4}
cN(|T|) \sim |T!|^{\del} e^{a|T|}.
\eeq
The numerical results they report are in total agreement with earlier
simulations in the case where we can check ($\k_0=0$ in the notation of
\cite{ckr}, see later)
and the purpose of this short note is to make clear that
there are other, in our opinion more plausible, interpretations of the
data, which are compatible with the exponential bound \rf{*3}. In addition
we have used the opportunity  to extend the simulations to somewhat larger
volume.

\vspace{24pt}

\noi{\large \bf 2. Results}

\vspace{12pt}

\noi In $4d$ the gravity model \rf{*1} can be written as \cite{aj}
\beq{*5}
Z = \sum_{T(S^4)} \frac{1}{C_T} e^{-k_4 N_4(T) +k_2 N_2(T)}
\eeq
where $N_4(T)$ and $N_2(T)$ denote the number of 4-simplexes and triangles
in the triangulation $T$. According to Regge calculus $k_2 \sim 1/G_0$, the
bare gravitational constant. If we have a regular triangulation of $S^4$ the
number of vertices $N_0(T)= 2-N_4(T)+N_2(T)/2$. This allows us to write
\beq{*6}
Z \sim  \sum_{T(S^4)} \frac{1}{C_T} e^{-\k_4 N_4(T) +\k_0 N_0(T)},
{}~~~~~~~\k_0 = 2 k_4,~~~\k_4 = k_4-2k_2,
\eeq
in order to make contact with the notation used in \cite{ckr}.
If we {\it assume} \rf{*3} there will for each value of
$\k_0$ be a critical value, $\k_4^c(\k_0)$ of $\k_4$,
such that the partition function
is divergent for $\k_4 < \k_4^c(\k_0)$ and
convergent for $\k_4 > \k_4^c(\k_0)$. The infinite volume limit
of the system is obtained by letting $\k_4$
approach $\k_4^c(\k_0)$ from above.
Whether this infinite volume limit has an interesting {\it continuum}
interpretation should then be investigated for various values of
$\k_0$. A second order phase transition in geometry might indicate
long range fluctuations and thereby continuum physics.
Details of how to determine
$\k_4^c(\k_0)$ can be found in \cite{ajk}, the only point
we want  to emphasize here is that one determines a pseudo-critical value
$\k_4^c(\k_0,N_4)$ by analyzing fluctuations in size around some
chosen $N_4$. By changing $N_4$ this pseudo-critical point will
change and in the limit $N_4\to \infty$ it should converge to
$\k_4^c(\k_0)$, again provided \rf{*3} is valid. In case \rf{*4} should be
valid there will be no convergence for $N_4 \to \infty$.

In order to discuss possible options, let us for simplicity choose $\k_0=0$.
We are then directly testing $\cN (N_4)$. Let us assume that the number
of configurations grows exponentially. There will be subleading corrections.
A reasonable trial ansatz for large $N_4$ is
\beq{*8}
\log (\cN(N_4))  \sim \k_4^cN_4  - \tilde{c} N_4^\b- \cdots,~~~~~~\b < 1.
\eeq
This is known to be true in $2d$ gravity where $\b=0$ (and the power
is replaced by a logarithm) and (to the extent one can trust
numerical simulations) in $3d$ where $\b \approx 2/3$.
This subleading behavior will reflect itself in the determination
of $\k_4^c(N_4)$ which will be given by
\beq{*9}
\k_4^c(N_4)= \k_4^c -  \frac{c}{N_4^{\a}} -\cdots,~~~~~~\a=1-\b.
\eeq
If we instead assume that \rf{*4} is valid we will get
\beq{*10}
\k_4^c (N_4) = a + \del \log N_4.
\eeq
Let us first remark that in both cases $\k_4^c(N_4)$ is an increasing
function of $N_4$ and in case $\a$ and $\del$ are small
it might be difficult to distinguish \rf{*9} and \rf{*10}
when only a limiting range of $N_4$ is considered. Secondly
one should be aware that both \rf{*9} and \rf{*10} are asymptotic
expressions which might only be valid for large $N_4$. Thirdly,
\rf{*9} and \rf{*10} are valid also for $\k_0 \neq 0$, the only difference
being that $\k_4^c$, $\a$ and $a$ might depend on $\k_0$, {\it while $\del$
is independent of $\k_0$} since  \rf{*2} implies that various
actions $S(T)$ can only lead to different exponential corrections.
$(N_4 !)^\del$ is entirely an entropy factor coming from the
number of triangulations.

In fig. 1 we show
$\k_4^c(N_4,\k_0)$ as presented in \cite{ckr} for three values
of $\k_0$. The
curves are just linear interpolations between the data points.
We see a clear deviation from the functional form
\rf{*9} or \rf{*10} for $N_4 < 4000$ in all three cases. Since
this deviation is qualitatively of the same form in the three cases
it is natural to assume that there are finite size effects not
compatible with \rf{*9} or \rf{*10} for $N < 4000$. In any case
it seems not reasonable to determine $\del$ from $N_4 < 4000$
since the functional form here shows a strong dependence on $\k_0$
(as also remarked in \cite{ckr}) and $\del$ is independent of $\k_0$.
Let us now concentrate on the case $\k_0 =0$ where we can directly
compare also with our own old (and new) results. In fig. 2 we
show  $\k_4^c(N_4)$ for $\k_0=0$ and $N_4 \geq 4000$.
When data points can be compared directly
there is agreement within error bars.
The (dashed) straight line
is the fit \rf{*10} found in \cite{ckr}.
The fully drawn line is a fit to \rf{*9} with $\a=1/4$ (the reason for this
choice will be discussed below). In both cases there are two free
parameters in the fit, $a,~\del$ and $\k^c_4,~c$, respectively.
For the  non-trivial parameters, i.e. $c$ and $\del$ we have
$c \approx 1.23$ and $\del \approx 0.03$.

\vspace{24pt}

\noi{\large \bf 3. Conclusions}

\vspace{12pt}

\noi
As is clear from fig. 2 the numerical simulations so far
are perfectly consistent with an exponential bound on the entropy
of triangulations of $S^4$.
While it is possible that one can still make \rf{*10} compatible with
fig. 2 even if we include the new data point corresponding to
$N_4=64000$, we feel that a fit of the form \rf{*9} is more
natural for $N > 4000$ for several reasons.
First it is hard to understand why there should be
a factorial growth $(N!)^\del$ with an extremely small
exponent $\del \sim 0.03$. As mentioned above the origin of such
a term is purely combinatorial and {\it if} there was a factorial
growth  one would a priori expect $\del \sim 1$. Secondly it is
difficult for us to understand why the situation for $S^4$ should
differ from that of $S^3$. Rather, since there is also an exponential bound
in $2d$ and $3d$ it is natural to conjecture that there is an exponential
bound on the number of triangulations of $S^d$ for any $d$.

It is interesting
to note that data are compatible\footnote{We should note
that $\a$ is not very well {\it determined} from a fit to the
data shown in fig. 2. Any $\a$ in the range $0.2 < \a < 0.3$ will give
a reasonable fit.} with the choice $\a=1/4$.
This means that the exponent $\b$ in \rf{*8} is 3/4 and if we
tentatively write $N_4 = L^4$ eq. \rf{*8} reads:
\beq{*11}
\log \cN(L^4) = \k_4^c L^4 -c L^3 -\cdots.
\eeq
In $3d$ we found \cite{av} $\b \approx 2/3$ and
introducing in $3d$ the notation
$N_3= L^3$ one has the analogue of \rf{*11}:
\beq{*12}
\log \cN(L^3) = \k_3^c L^3 -c L^2 -\cdots.
\eeq
This kind of behavior might be related to the fact that
the generic triangulation for $\k_0=0$, both in $3d$ and $4d$
is very singular. There are vertices of very high order (almost
comparable with $|T|$) and the $d-1$ dimensional boundary of the
unit ball surrounding such a vertex could maybe act as  "pseudo-boundary"
of the closed $d$-dimensional manifold.  It would explain the
boundary-like terms in \rf{*11} and \rf{*12}. When $\k_0$ is increased
(decreasing bare gravitational constant) the vertices of very high order
disappear. At the same time the value of $\b$ seems to decrease, maybe
approaching zero at the observed phase transition\footnote{The value of
$\b$ (or $\a$) is not very well determined from the fits to
$\k_4^c(N_4)$ by the present data for $\k_0$ near or above the
phase transition point. In $3d$ the situation is
better and it seems as if $\b \approx 0$ at the transition
\cite{abkv,av1}. The study
of baby-universe distributions \cite{ajjk} also supports this picture.}.
In addition it would also
explain why the corresponding term $c L$ is  absent in $2d$.
The distribution of high-order vertices in $2d$ can be calculated
analytically and the probability of having such vertices
falls off exponentially with the order of the vertex.

\newpage

%\addtolength{\baselineskip}{-0.20\baselineskip}

\newpage

\begin{center}
{\large \bf Figure caption}
\end{center}

\begin{itemize}
\item[{\bf Fig. 1}] The data points taken from \cite{ckr}. The plot is
$\log N_4$ versus $\k_4^c(N_4,\k_0)$ for $\k_0=0$ (lower curve),
$\k_0 = 0.5$ (middle curve) and $\k_0 = 1.0$ (upper curve).
\vspace{12pt}

\item[{\bf Fig. 2}] $\log N_4$ versus $\k_4^c(N_4,\k_0)$ for
$\k_0=0$. Circles indicate the data points from
the plot in \cite{ckr}, the size of the circles
being approximately the error bars.
The dots (with error bars)  indicate our own data.
We see that there is perfect agreement. The dashed straight line is
the fit to \rf{*10} as presented in \cite{ckr}. The fully drawn
line is the fit to \rf{*9} with $\a=1/4$.
\end{itemize}


\begin{thebibliography}{99}
\bibitem{aj}J. Ambj\o rn and J. Jurkiewicz, Phys.Lett B278 (1992) 42.
\bibitem{am}M.E. Agishtein and A.A. Migdal, Mod. Phys. Lett. A7 (1992) 1039.
\bibitem{adf}J. Ambj\o rn, B. Durhuus and J. Fr\"{o}hlich, Nucl. Phys. B 257
 (1985) 433; Nucl.Phys. B270 (1986) 457; Nucl. Phys.  B275 (1986) 161.
\bibitem{david1}F. David, Nucl. Phys. B 257 (1985) 45;
 Nucl. Phys. B257 (1985) 543.
\bibitem{kkm}V. A. Kazakov, I. K. Kostov and A. A. Migdal, Phys. Lett. 157B
 (1985) 295; Nucl.Phys. B275 (1986) 641.
\bibitem{adj}J. Ambj\o rn, B. Durhuus and T. Jonsson,
Mod.Phys.Lett A6 (1991) 1133.
\bibitem{av}J. Ambj\o rn and S. Varsted, Phys. Lett (B226), (1991) 285.
\bibitem{sakura}N. Sakura, Mod.Phys.Lett. A6 (1991) 2613.
\bibitem{gross}N. Godfrey and M. Gross, Phys.Rev. D43 (1991) R1749.
\bibitem{am1}M.E. Agishtein and A.A. Migdal, Mod. Phys. Lett. A6 (1991) 1863.
\bibitem{bk}B. Boulatov and A. Krzywicki, Mod.Phys.Lett A6 (1991) 3005.
\bibitem{abkv}J. Ambj\o rn, D. Boulatov, A. Krzywicki and S. Varsted,
Phys.Lett B276 (1992) 432.
\bibitem{av1}J. Ambj\o rn and S. Varsted, Nucl.Phys. B373 (1992) 557.
\bibitem{tutte}W.T. Tutte, Canad. J. Math. 14 (1962) 21.
\bibitem{ckr}S. Catterall, J. Kogut and R. Renken, {\it On the absence
of an exponential bound in four dimensional simplicial gravity},
CERN-TH-7197/94.
\bibitem{ajk}J. Ambj\o rn, J. Jurkiewicz and C.F. Kristjansen,
Nucl.Phys. B393 (1993) 601.
\bibitem{ajjk}J. Ambj\o rn, S. jain, J. Jurkiewicz and C.F.
Kristjansen, Phys.Lett. B305 (1993) 208.
\end{thebibliography}
\end{document}